\documentclass[amsmath,amssymb,showpacs,draft,preprint,floatfix]{revtex4}
\usepackage{graphicx}
\usepackage{latexsym}
\usepackage{amsmath}
\usepackage{amsfonts}
\usepackage{amssymb}
\begin{document}

\title{The Glauber-Sudarshan and Kirkwood-Rihaczec functions}
\author{ H\'ector Manuel Moya-Cessa  }
\affiliation{Instituto Nacional de Astrof\'{\i}sica, \'Optica y
Electr\'onica, Calle Luis Enrique Erro No. 1, 72840 Santa
Mar\'{\i}a Tonantzintla, Pue.}
\begin{abstract}
 It is shown how to write the Kirkwood-Rihaczec 
quasiprobability distribution as an expectation value of the vacuum state. We do this, by writing
the position eigenstates as a "displacment" of the vacumm.
We also give a relation between the Glauber-Sudarshan and Kirkwood-Rihaczec 
quasiprobability distributions.
\end{abstract}
 \maketitle
\section{Introduction}\label{intro}
Quasiprobability distribution functions are widely used in quantum mechanics \cite{Hillery,Schleich} and optical physics
\cite{Wolf}. One of the best known
quasiprobability distribution functions is the  Wigner function
\cite{Hillery,Wigner32} with applications in reconstruction of
signals \cite{Wolf}  in the classical
regime and reconstruction  of quantum states of different systems
such as ions \cite{Wineland} or quantized fields
\cite{Haroche,Moya2} in the quantum regime. Quasiprobability distribution functions are also useful to identify non-classical states of light \cite{6,Moya3}. In this
contribution we would like to re-introduce a lesser known
quasiprobability distribution function, namely the
Kirkwood-Rihaczek function
\cite{Kirkwood,Rihaczek,Praxmayer1,Praxmayer2}, and use its relation to the Wigner function to
show how it may be related to the Glauber-Sudarshan $P$-function.
This may be done as we express the Kirkwood-Rihaczec function as an expectation value in term of the vacuum.

\subsection{Wigner function}
We start by introducing the Wigner function, probably the best
known. It may be written in two forms: series representation (see
for instance \cite{series}), and integral representation
\begin{equation}
W(q,p) = \frac{1}{2\pi}\int du e^{iup}\langle q+\frac{u}{2}|\rho
|q-\frac{u}{2}\rangle , \label{wigner}
\end{equation} 
where $\rho$ is the density matrix. In
1932, Wigner introduced this function $W(q,p)$, known now as
 his distribution function \cite{Wigner32,Hillery} and contains
complete information about the state of the system as the density matrix for a pure state is given by
$\rho=|
\psi \rangle \langle\psi|$.

The Wigner function may be written also as in terms of the (double) Fourier
transform of the characteristic function
\begin{equation}
W(\alpha )= \frac{1}{4\pi ^{2}}\int \exp (\alpha \beta ^{\ast
}-\alpha ^{\ast }\beta )C(\beta )d^{2}\beta, \label{wigcar}
\end{equation}
with $\alpha=(q+ip)/\sqrt{2}$ and where $C(\beta )$ in terms of
annihilation and creation operators is given by
\begin{equation}
C(\beta )=Tr\{{\rho}\exp (\beta {a}^{\dagger }-\beta ^{\ast }{a}%
)\}, \label{caracteristic}
\end{equation}
also known as ambiguity function in classical optics
\cite{Stenholm}. $a$ and $a^{\dagger}$ are the annihilation and creation operators for the harmonic oscillator.

\subsection{Cohen-class distribution functions}
A function of the Cohen class is described by the general formula
\cite{Cohen}
\begin{equation}
W_C=\frac{1}{2\pi}\int\int\int\phi(y+\frac{1}{2}x')
\phi(y-\frac{1}{2}x')k(x,u,x',u')e^{-i(ux'-u'x+u'y)}dxdx'du'
\end{equation}
and the choice of the kernel $k(x,u,x',u')$ selects one particular
function of the Cohen class. The Wigner function, for instance
arises for $k(x,u,x',u')=1$, whereas the ambiguity function is
obtained for $k(x,u,x',u')2\pi\delta(x-x')\delta(u-u')$.

\section{The Kirkwood-Rihaczek quasidistribution function}
Now we turn our attention to a lesser known distribution, the
Kirkwood-Rihaczek function, that may be written using the notation
above as \cite{Lee}
\begin{equation} K\left( \beta \right)
=\int d^{2}\alpha e^{\beta \alpha ^{\ast }-\beta ^{\ast }\alpha
}e^{\frac{\alpha ^{2}-\alpha ^{\ast 2}}{4}}C\left( \alpha \right)
\label{kirk} ,
\end{equation}
may also be expressed as the double Fourier transform
\begin{equation} K\left( q,p \right)
=\int du dv e^{-iup} e^{ivq} Tr\{\rho e^{iv{q}}e^{iu{p}}\},
\end{equation}
and the trace is to be taken in the form
\begin{equation} 
Tr\{\rho A\}=\langle\psi|A|\psi\rangle=\int_{-\infty}^{\infty}dq\langle q|\psi\rangle\langle \psi|A|q\rangle.
\end{equation}

We will now do an analysis similar to the one done in reference \cite{series}. We relate the
Kirkwood-Rihaczek function to the Wigner function by using
(\ref{kirk}), via the following exponential of derivatives
\begin{equation}
K\left(
\beta \right)=e^{-\frac{1}{4}\frac{\partial ^{2}}{\partial ^{2}\beta }}e^{\frac{1}{4}\frac{%
\partial ^{2}}{\partial ^{2}\beta ^{\ast }}}W\left( \beta \right).
\end{equation}
In the above equation we will use a non-integral expression for the Wigner function
\cite{series}
\begin{equation}
W\left( \beta \right) =Tr\left[ \left( -1\right) ^{{a^{\dagger}a}}{D}%
^{\dagger }\left( \beta \right) {\rho }{D}\left(
\beta \right) \right],
\end{equation}
with ${D}\left( \beta \right)=e^{\beta
a^{\dagger}-\beta^*a}$, the so-called Glauber displacement
operator. Rearranging the displacement operators and the parity operator, we obtain
\begin{equation}
W\left( \beta \right) =Tr\left[ \left( -1\right)
^{{a^{\dagger}a}}{\rho }{D}\left( 2\beta \right)
\right],
\end{equation}
where we have used the trace property $Tr(AB)=Tr(BA)$ and the
following identities$\left( -1\right)
^{{a^{\dagger}a}}{D}^{\dagger }\left( \beta \right)
={D}\left( \beta \right) \left( -1\right) ^{{a^{\dagger}a}}$.

Now we use the factorized form of the Glauber displacement
operator \cite{Glauber} $ {D}\left( 2\beta \right)
=e^{-2\left| \beta \right| ^{2}}e^{2\beta
{a}^{\dagger }}e^{-2\beta ^{\ast }{a}}$ to obtain
\begin{equation}
W\left( \beta \right) =Tr\left[ \left( -1\right)
^{{a^{\dagger}a}}{\rho }e^{-2\left| \beta \right|
^{2}}e^{2\beta {a}^{\dagger }}e^{-2\beta ^{\ast
}{a}}\right].
\end{equation}
Therefore, we have that the Kirkwood-Rihaczek function may be
written as
\begin{eqnarray}
\nonumber
K\left( \beta ,\beta ^{\ast }\right) &=&e^{-\frac{1}{4}\frac{\partial ^{2}}{%
\partial ^{2}\beta }}e^{\frac{1}{4}\frac{\partial ^{2}}{\partial ^{2}\beta
^{\ast }}}W\left( \beta ,\beta ^{\ast }\right) \\
&=&Tr\left[ \left( -1\right) ^{{a^{\dagger}a}}{\rho }e^{-\frac{1}{4}\frac{%
\partial ^{2}}{\partial ^{2}\beta }}e^{\frac{1}{4}\frac{\partial ^{2}}{%
\partial ^{2}\beta ^{\ast }}}{D}\left( 2\beta \right)
\right].
\end{eqnarray}
The calculation of the exponential of derivatives of the Glauber
operator will be  tedious but straightforward. We note that
\begin{equation}
e^{\frac{1}{4}\frac{\partial ^{2}}{\partial ^{2}\beta ^{\ast }}}{D}%
\left( 2\beta \right) =e^{-\beta ^{2}}e^{2\beta \left(
{a}^{\dagger
}+{a}-\beta ^{\ast }\right) }e^{{a}^{2}}e^{-2\beta ^{\ast }%
{a}},
\end{equation}
and by using the expression for the generating function for Hermite polynomials \cite{Arfken}
\begin{equation} \label{generating}
e^{-t^{2}+2tx}=\sum_{k=0}^{\infty }H_{k}\left( x\right)
\frac{t^{k}}{k!}
\end{equation}%
we can express the above equation as
\begin{equation}
e^{\frac{1}{4}\frac{\partial ^{2}}{\partial ^{2}\beta ^{\ast }}}{D}%
\left( 2\beta \right) =\sum_{k=0}^{\infty }H_{k}\left(
{a}^{\dagger
}+{a}-\beta ^{\ast }\right) \frac{\beta ^{k}}{k!}e^{{a}%
^{2}}e^{-2\beta ^{\ast }{a}}.
\end{equation}%
 From the above equation, it is
easy to note that
\begin{equation}
\frac{\partial ^{2n}}{\partial \beta ^{2n}}\sum_{k=0}^{\infty
}H_{k}\left( x\right) \frac{\beta ^{k}}{k!}=\sum_{k=0}^{\infty
}H_{k+2n}\left( x\right) \frac{\beta ^{k}}{k!}
\end{equation}
such that
\begin{equation}
e^{-\frac{1}{4}\frac{\partial ^{2}}{\partial ^{2}\beta }}e^{\frac{1}{4}\frac{%
\partial ^{2}}{\partial ^{2}\beta ^{\ast }}}{D}\left( 2\beta \right)
=\sum_{n=0}^{\infty }\sum_{k=0}^{\infty }\frac{\left(
-\frac{1}{4}\right) ^{n}}{n!}H_{k+2n}\left( {a}^{\dagger
}+{a}-\beta ^{\ast }\right) \frac{\left( \beta \right)
^{k}}{k!}e^{{a}^{2}}e^{-2\beta ^{\ast }{a}}.
\end{equation}
Now we use the integral form of the Hermite polynomials
\cite{Arfken}
\begin{equation}
H_{p}\left( x\right) =\frac{2^{p}}{\sqrt{\pi
}}\int\limits_{-\infty }^{\infty }\left( x+it\right)
^{p}e^{-t^{2}}dt
\end{equation}
to obtain
\begin{eqnarray}\nonumber
K\left( \beta ,\beta ^{\ast }\right)& =&\frac{e^{-\beta ^{\ast
2}}e^{-2\beta \beta ^{\ast }}}{\sqrt{\pi }}\int\limits_{-\infty
}^{\infty }dx\int\limits_{-\infty }^{\infty }dt e^{\left(
-2\sqrt{2}x+2\beta ^{\ast }+2\beta \right)
it}e^{-2x^{2}}e^{2\sqrt{2}x\left( \beta ^{\ast }+\beta
\right) }\\
&\times &
\left\langle x\right| e^{{a}^{2}}e^{-2\beta ^{\ast }%
{a}}\left( -1\right) ^{{a^{\dagger}a}}{\rho
}\left| x\right\rangle
\end{eqnarray}
by using
\begin{equation}
\int\limits_{-\infty }^{\infty }e^{-iyt}dt=2\pi \delta \left(
y\right).
\end{equation}
If we take $y=2\sqrt{2}x-2\beta ^{\ast }-2\beta $ we have
\begin{eqnarray}\nonumber
K\left( \beta ,\beta ^{\ast }\right) &=&2\sqrt{\pi }e^{-\beta
^{\ast 2}}e^{-2\beta \beta ^{\ast }}\int\limits_{-\infty }^{\infty
}dx\delta
\left( 2\sqrt{2}x-2\beta ^{\ast }-2\beta \right) e^{-2x^{2}}e^{2\sqrt{2}%
x\left( \beta ^{\ast }+\beta \right) } \\
&&\times \left\langle x\right| e^{{a}^{2}}e^{-2\beta ^{\ast }%
{a}}\left( -1\right) ^{{a^{\dagger}a}}{\rho
}\left| x\right\rangle .
\end{eqnarray}
Making use of the identity $\delta \left( \alpha x\right) =\frac{\delta \left( x\right) }{%
\left| \alpha \right| }$ we finally obtain

\begin{equation} \label{kirkx}
K\left( \beta ,\beta ^{\ast }\right) =\sqrt{\frac{\pi}{2}}e^{\beta ^{2}-\beta ^{\ast 2}}\left\langle X
\right| e^{\left( {a}-\beta ^{\ast }\right)^{2}\left( -1\right) ^{{a^{\dagger}a}}\rho }\left| X\right\rangle
\end{equation}%
with $\beta=\frac{X+iY}{\sqrt{2}}$.

The position eigenstate $\left| X\right\rangle$ may be written as
\begin{equation}
\left| X\right\rangle = \sum_{n=0}^{\infty} \psi_n(X)|n\rangle
\end{equation}%
with $\psi_n(X)=\frac{e^{-X^2/2}H_n(X)}{\sqrt{2^n\sqrt{\pi}n!}}$
such that the position eigenstate may we re-written as
\begin{equation}
\left| X\right\rangle =\frac{ e^{-X^2/2}}{\pi^{1/4}}\sum_{n=0}^{\infty}\frac{H_n(X)}{{2^{n/2}}n!}a^{\dagger n}|0\rangle
\end{equation}
that may be added via the generating function for Hermite polynomials(\ref{generating}) to give
\begin{equation}
\left| X\right\rangle =\frac{ e^{-X^2/2}}{\pi^{1/4}}e^{-\frac{a^{\dagger 2}}{2}+\sqrt{2}a^{\dagger}X}|0\rangle.
\end{equation}
In the above equation the application of an operator to the vacuum produces the position eigenstate.

By using the above expression in equation (\ref{kirkx}) we obtain

\begin{equation} \label{kirk-0}
K\left( \beta ,\beta ^{\ast }\right) ={\frac{e^{\beta ^{2}-X^2}
}{\sqrt{2}}}\left\langle 0
\right| e^{ \frac{{a}^2}{2}-\sqrt{2}iYa}\left( -1\right) ^{{a^{\dagger}a}}\rho e^{-\frac{a^{\dagger 2}}{2}+\sqrt{2}a^{\dagger}X}\left| 0\right\rangle,
\end{equation}%
or
\begin{equation} \label{kirk-00}
K\left( \beta ,\beta ^{\ast }\right) ={\frac{e^{\beta ^{2}-X^2}
}{\sqrt{2}}}\left\langle 0
\right| e^{ \frac{{a}^2}{2}+\sqrt{2}iYa}\rho e^{-\frac{a^{\dagger 2}}{2}+\sqrt{2}a^{\dagger}X}\left| 0\right\rangle,
\end{equation}%
that may be finally written as an expectation value in terms of coherent states
\begin{equation} \label{kirk-coh}
K( \beta ,\beta ^{\ast } )=\frac{e^{\beta^2+Y^2}
}{\sqrt{2}}\langle -\sqrt{2}iY
| e^{ \frac{{a}^2}{2}}\rho e^{-\frac{a^{\dagger 2}}{2}}| \sqrt{2}X \rangle.
\end{equation}

We can relate the Kirkwood function to the Glauber-Sudarshan $P$-function \cite{Glauber,Sudarshan} by
using the relation $\rho=\int d^2\alpha P(\alpha)|\alpha \rangle\langle \alpha |$, i.e.,
\begin{equation} \label{kirk-P}
K( \beta ,\beta ^{\ast } )=\frac{e^{\beta^2+Y^2}
}{\sqrt{2}}\int d^2\alpha P(\alpha)e^{\frac{\alpha^2-Ê\alpha^{* 2}}{2}}\langle -\sqrt{2}|Y
|\alpha \rangle\langle \alpha | \sqrt{2}X \rangle,
\end{equation}
or
\begin{equation} \label{kirk-P-final}
K( \beta ,\beta ^{\ast } )=\frac{e^{iXY}
}{\sqrt{2}}\int d^2\alpha P(\alpha)e^{\frac{\alpha^2-Ê\alpha^{* 2}}{2}-|\alpha|^2}e^{\sqrt{2}(X\alpha^*-iY\alpha)}.
\end{equation}

Therefore we have written the Kirkwood-Rihaczek
function as an expectation value in terms of the vacuum state, just as the $Q$-function may be written as a coherent states expectation value,  the Wigner and Glauber-Sudarshan functions in
terms of a series of displaced number states expectation values \cite{series}, and relate it to the Glauber-Sudarshan $P$-function.
\section{Conclusions}
We have written the position eigenstates as a "displacement" of the vacuum state, which has allowed us to use a
former expression for the Kirkwood-Rihaczec distribution function to write it as an expectation value in terms of the vacuum state.
This made easy to relate this function to the Glauber-Sudarshan $P$-function \cite{Puri}.

\end{document}